\documentclass{jetpl}
\twocolumn
\usepackage[usenames,dvipsnames]{xcolor} 
\usepackage[
    colorlinks = true,
    urlcolor = MidnightBlue,
    linkcolor = Maroon,
   citecolor = Maroon
]{hyperref}

\hypersetup{
    pdfauthor = {A.V. Glushkov, L.T. Ksenofontov, K.G. Lebedev, A.Saburov},
    pdftitle = {Estimation of the composition of ultra-high-energy cosmic rays using the muon correlation method based on Yakutsk EAS array data},
    pdfsubject = {Mass composition of ultra-high energy cosmic rays},
    pdfkeywords = {extensive air showers, muon component, mass composition, Yakutsk array}
}

\usepackage{cite}

\graphicspath{{figures/}}

\lat

\newcommand{\doi}[2] {%
    \href{https://doi.org/#2}{#1}%
}

\newcommand{\arXiv}[2] {%
    arXiv:\href{https://arXiv.org/abs/#1}{#1} [#2]%
}

\newcommand{\arXivOld}[2] {%
    arXiv:#2/\href{https://arXiv.org/abs/#2/#1}{#1}%
}

\newcommand{\fign}[1]{Fig.~\ref{#1}}
\newcommand{\eqn}[1]{(\ref{#1})}

\newcommand{\mean}[1]{\left<{#1}\right>}

\newcommand{\E}{E_0}

\renewcommand{\lg}{\log_{10}}

\newcommand{\Nmax}{N_{\text{max}}}
\newcommand{\xmax}{x_{\text{max}}}
\newcommand{\xmaxA}{x_{\text{max}}^A}
\newcommand{\xmaxP}{x_{\text{max}}^p}
\newcommand{\xmaxFe}{x_{\text{max}}^{\text{Fe}}}
\newcommand{\xmaxExp}{x_{\text{max}}^{\text{exp}}}

\newcommand{\AFe}{A_{\text{Fe}}}

\newcommand{\zA}{z_A^{\text{exp}}}
\newcommand{\rhoMu}[1]{\rho_{\mu}^{#1}}

\newcommand{\Ssoo}{S_{600}}
\newcommand{\SsExp}{S_{600}^{\text{exp}}}
\newcommand{\SsSim}{S_{600}^{\text{sim}}}

\newcommand{\Smu}{S_{\mu}}
\newcommand{\Smus}{S_{\mu, 600}}
\newcommand{\SmuSim}{S_{\mu}^{\text{sim}}}
\newcommand{\SmuExp}{S_{\mu}^{\text{exp}}}
\newcommand{\SmusSim}{S_{\mu, 600}^{\text{sim}}}
\newcommand{\SmusExp}{S_{\mu, 600}^{\text{exp}}}
\newcommand{\SmutSim}{S_{\mu, 1000}^{\text{sim}}}
\newcommand{\SmutExp}{S_{\mu, 1000}^{\text{exp}}}

\newcommand{\rM}{r_{\text{M}}}

\newcommand{\sqrm}{m$^2$}     
\newcommand{\depth}{g/cm$^2$} 
\newcommand{\pdens}{m$^{-2}$} 
\newcommand{\degr}{^{\circ}}  

\newcommand{\qgs}{{\sc qgsj}et01}
\newcommand{\qgsii}{{\sc qgsj}et-{\sc ii}.04}

\newcommand{\eposlhc}{{\sc epos-lhc}}

\newcommand{\corsika}{{\sc corsika}}

\title{Estimation of the composition of ultra-high energy cosmic rays using the muon correlation method based on Yakutsk EAS array data}

\rtitle{Estimation of the composition of ultra-high-energy\ldots}

\sodtitle{Estimation of the composition of ultra-high-energy cosmic rays using the muon correlation method based on Yakutsk EAS array data}

\author{A.\,V.\,Glushkov$^{*}$\/\thanks{e-mail: glushkov@ikfia.ysn.ru},
L.\,T.\,Ksenofontov$^*$, K.\,G.\,Lebedev$^*$, A.\,V.\,Saburov}

\rauthor{A.\,V.\,Glushkov, L.\,T.\,Ksenofontov, K.\,G.\,Lebedev, A.\,V.\,Saburov}

\sodauthor{Glushkov, Ksenofontov, Lebedev, Saburov}

\address{$^*$Yu.\,G.\,Shafer Institute of Cosmophysical Research and Aeronomy SB RAS\\
677027 Yakutsk, Russia}

\abstract{In this article a new method is proposed for estimating the mass composition of cosmic rays in individual events with energies above $1.25 \times 10^{19}$~eV. It is based on a joint analysis of experimental data and simulation results obtained using the \qgsii{} model for muons with threshold energy $E_{\mu} = 1.0 \times \cos\theta$~GeV in air showers with zenith angles up to 60 degrees. The data from ground-based and underground scintillation detectors of the Yakutsk EAS array were used. Separate groups of nuclei and other primary particles were found.}

\begin{document}

\maketitle

\section{Introduction}

Cosmic ray (CR) mass composition in ultra-high energy range (above $\sim 10^{17}$~eV) so far is studied only in general terms. Information about it is often contradictory. Atmospheric depth $\xmax$~--- at which the maximum number of particles ($\Nmax$) is achieved during the extensive air shower (EAS) development~--- is a standard parameter for obtaining information about the CR composition. Simulations demonstrate that depending on the mass number $A$ of primary particles, the process of air shower longitudinal development, taking into account fluctuations, leads to certain values of the mean depth $\mean{\xmaxA}$ and its dispersion $\delta(\xmaxA)$. The $\mean{\xmax}$ value is connected to conventional mean atomic number $\mean{A}$ of primary particles via a simple ratio, which follows from the nucleon superposition principle~\cite{b:1}:

\begin{equation}
    \mean{\ln{A}} = 
    \frac{\
        \mean{\xmaxA} - \mean{\xmaxP}
    }{
        \mean{\xmaxFe} - \mean{\xmaxP}
    }
    \times \ln\AFe\text{,}
    \label{eq:1}
\end{equation}
where values for primary protons ($p$) and iron nuclei (Fe) were obtained using one or another hadron interaction model. In experiment the $\mean{\ln{A}}$ value is determined by substituting the $\mean{\xmaxA}$ with measured value $\mean{\xmaxExp}$. This technique is successfully utilized at Auger~\cite{b:2} and Telescope Array (TA)~\cite{b:3} experiments for individual EAS events with registered fluorescent light emission.

Simulations have revealed that expression~\eqn{eq:1} is applicable to other EAS parameters that are sensitive the $\xmax$ value. This is especially true for muon component, which is actively studied in many experiments in the wide range of primary energy. In this case one has to deal with the following relations:

\begin{gather}
    \mean{\ln{A}}_{\text{exp}} = \mean{\zA} \cdot \ln{\AFe}\text{,}
    \label{eq:2} \\
    \mean{\zA} = \frac{
        \ln\mean{\rhoMu{\rm exp.}} - \ln\mean{\rhoMu{p}}
    }{
        \ln\mean{\rhoMu{\rm Fe}} - \ln\mean{\rhoMu{p}}
    }\text{,}
    \label{eq:3}
\end{gather}
where $\mean{\rhoMu{\rm exp.}}$ is the muon density registered in experiment and $\mean{\rhoMu{p}}$ and $\mean{\rhoMu{\rm Fe}}$ are densities obtained for primary protons ($p$) and iron nuclei (Fe) using full simulation of the measurement process with a real detector. The value \eqn{eq:3} has been heavily discussed recently due to existing fundamental disagreements~\cite{b:4, b:5, b:6, b:7}.

At the Yakutsk array the muon component of EAS has been registered since the very beginning of its operation in 1974. To date, a significant experimental material has been accumulated, which is episodically analyzed as the notion of the nature of CR evolves. Below, a new method is considered for estimating the CR mass composition based on these data. It relies on formulas \eqn{eq:2} and \eqn{eq:3} in relation to the physical picture of EAS development.

\section{CR mass composition}

\subsection{General formulation of the problem}

The Yakutsk EAS array stands out from other similar instruments by its complex design: it simultaneously measures charged and electromagnetic particles with $2 \times 2$-\sqrm{} surface scintillation detectors (SD), muons with energy above $1 \times \sec\theta$~GeV with similar ground shielded detectors (MD) with an area $20-36$~\sqrm{} and EAS Cherenkov light emission (CLE). In the work~\cite{b:8} lateral distribution functions (LDFs) of SD and MD responses and zenith-angular dependencies of the corresponding densities at axis distance $r = 600$~m were studied in showers with $\E = 10^{19}$~eV and zenith angles $\theta \le 60\degr$. Experimentally measured values were compared to estimations obtained within frameworks of hadron interaction models \qgs~\cite{b:9} and \qgsii~\cite{b:10} using the \corsika{} code~\cite{b:11}. The details of the SD and MD response calculation are given in~\cite{b:7, b:8}. The whole considered data set indicates a certain agreement between experiment and theory. In~\cite{b:8} it is stated that the probable CR composition is close to protons with possible fraction of primary photons about $9$\%. This conclusion was based on the analysis of mean LDFs in separate groups of showers with zenith-angular directions $\mean{\cos\theta} = 0.95, 0.90, 0.85, 0.80, 0.75, 0.65$ and $0.55$. Here we present the results of a further study of the CR mass composition with energy above $1.25 \times 10^{19}$~eV in individual events. Let's reiterate on some major points of the procedure for processing of registered events adopted at the Yakutsk array, which we will need later.

\subsection{Energy estimation}

CLE contains information about approximately $80$\% of primary energy $E$ dispersed by a shower in the atmosphere and provides the possibility to estimate the $E$ value with the use of calorimetric method~\cite{b:12, b:13, b:14, b:15, b:16}. The EAS energy was determined from following relations~\cite{b:12}:

\begin{gather}
    E = E_1 \times \Ssoo(0\degr)^B~\text{[eV],}
    \label{eq:4}\\
    \begin{aligned}
        & \Ssoo(0\degr) = \Ssoo(\theta) \times \\
        & \qquad\times
        \exp\left[
            \frac{
                (\sec\theta - 1) \times 1020
            }{
                \lambda
            }
        \right]~\text{[\pdens],}
        \label{eq:5}
    \end{aligned}\\
    \lambda = 400 \pm 45~\text{[\depth],}
    \label{eq:6}
\end{gather}
where $E_1 = (4.1 \pm 1.4) \times 10^{17}$~eV, $B = 0.97 \pm 0.04$. Numerically the $E_1$ proportional coefficient is equal to the energy of a vertical shower with density of the SD response measured at axis distance $600$~m $\Ssoo(0\degr) = 1$~\pdens. The experiment essentially measures primary energy in units of some reference EAS event. Later the parameters of the relation \eqn{eq:4} changed slightly: $E_1 = (4.8 \pm 1.6) \times 10^{17}$~eV, $B = 1.00 \pm 0.02$ and the expression for attenuation length~\eqn{eq:6} took the following form~\cite{b:13, b:14}:

\begin{align}
    \lambda = & (450 \pm 44) + (32 \pm 15) \times \nonumber \\
              & \qquad\times \log_{10}\Ssoo(0\degr)~\text{[\depth].}
    \label{eq:7}
\end{align}

In~\cite{b:15, b:16} we have reconsidered the energy calibration according to results of simulations performed with \corsika{} code. After that, the relations \eqn{eq:4}, \eqn{eq:5} and \eqn{eq:7} were finally adopted with parameters $E_1 = (3.76 \pm 0.30) \times 10^{17}$~eV and $B = 1.02 \pm 0.02$~\cite{b:16}. The error in formula \eqn{eq:4} was mainly conditioned by the precision of absolute calibration of CLE detectors and incorrect estimation of the atmospheric transparency~\cite{b:12}.

\subsection{Primary events reconstruction}

Arrival direction of a shower is reconstructed from relative delays of SD's firing using the approximation of flat shower front. The axis is located with the use of the Greisen-Linsley LDF approximation with parameters obtained at the Yakutsk array during the initial period of its operation~\cite{b:17}:

\begin{equation}
    S(r, \theta) = \Ssoo(\theta) \cdot \frac{600}{r} \cdot
    \left(
        \frac{600 + \rM}{r + \rM}
    \right)^{\beta(\theta) - 1}\text{,}
    \label{eq:8}
\end{equation}
where $\Ssoo(\theta)$ is the all-particle response density measured by scintillation SDs at axis distance $r = 600$~m and $\rM$ is the Moliere radius. It depends on atmospheric parameters $T$~[K] and $P$~[mb]~\cite{b:17, b:18}:

\begin{equation}
    \rM = \frac{7.5 \times 10^4}{P} \times \frac{T}{273}\text{.}
    \label{eq:9}
\end{equation}
Values of $T$ and $P$ are constantly monitored and recorded in the primary data bank at every triggering of the array. The annual mean value of the Moliere radius for Yakutsk is $\mean{\rM} \approx 70$~m. The structural parameter $\beta$ was determined in~\cite{b:17}:

\begin{align}
    \beta(\theta) &= 1.38 + 2.16 \times \cos\theta + \nonumber\\
                  &\qquad + 0.15 \times \log_{10}\Ssoo(\theta)\text{.}
    \label{eq:10}
\end{align}

Later it was established that the LDF approximation \eqn{eq:8} poorly describes experimental data for showers with $E \ge 10^{19}$~eV in the axis distance range $r \simeq 30-2000$~m, and a modified LDF approximation was introduced~\cite{b:19}:

\begin{gather}
    S(r, \theta) = \Ssoo(\theta) \cdot 
    \left(\frac{600}{r}\right)^2 \cdot
    \left(\frac{600 + \rM}{r + \rM}\right)^{\beta(\theta) - 2}
    \times \nonumber \\
    \times \left(
        \frac{600 + r_1}{r + r_1}
    \right)^{10}\text{,}
    \label{eq:11}
\end{gather}
where $r_1 = 10^4$~m. The $\beta(\theta)$ parameter in this case virtually does not change with the energy but depends on the zenith angle. The arrival direction, axis coordinates and the $\Ssoo(\theta)$ value are determined via $\chi^2$-minimizations during the preliminary processing of registered events after a diurnal cycle of the array operation.

The LDF of muon component in this work is approximated in the axis distance range $r \approx 300-2000$~m with the function:

\begin{align}
    &\Smu(r, \theta) = \Smus(\theta) \cdot
    \left(\frac{600}{r}\right)^{0.75} \times \nonumber \\
    & \qquad \times
    \left(\frac{600 + r_0}{r + r_0}\right)^{b_{\mu}(\theta) - 0.75}
    \cdot
    \left(\frac{600 + r_1}{r + r_1}\right)^9\text{,}
    \label{eq:12}
\end{align}
where $r_0 = 280$~m and $r_1 = 2000$~m. Absolute and daily calibrations of SDs and MDs were described in~\cite{b:7}.

\subsection{The muon correlation method}

Here we describe a method for estimating the CR composition in individual EAS events from muon component, which was called {\sl ``the muon correlation method''}. It consists in comparing the measured density of the MD response $\SmuExp(E, \theta, r)$ in a shower with energy $E$ and zenith angle $\theta$ at axis distance $r$ with the expected value $\SmuSim(E^*, \theta, r)$ obtained within the framework of the \qgsii{} model for primary proton with fixed energy $E^*$, the same zenith angle at the same axis distance. Simulations indicate that for registered events with energies $E \ge 10^{19.1}$~eV and $\theta \le 60\degr$, which were selected for further analysis, it is possible to determine the expected muon density from the relation:

\begin{equation}
    \SmuSim(E, \theta, r) = 
    \SmuSim(E^*, \theta, r) \cdot
    \frac{\SsExp(E, \theta)}{\SsSim(E^*, \theta)}\text{,}
    \label{eq:13}
\end{equation}
where $E^* = 3.16 \times 10^{19}$~eV. The $\SsSim(E^*, \theta)$ value reflects the zenith-angular dependency of the SD response density at $r = 600$~m. For a vertical shower it is connected via expression \eqn{eq:4} to the energy $E^*$, which is included in the simulation, and within $\approx 6$\% agrees with primary energy estimation obtained at the Yakutsk array~\cite{b:16}. The $\SsExp(E, \theta) = \Ssoo(\theta)$ parameter is estimated experimentally and is included in expression \eqn{eq:4}. Further we consider the correlation between densities $\SmuSim(E, \theta, r)$ at $r = 600$ and $1000$~m and the corresponding experimentally derived values.

\begin{figure}[!htb]
    \centering
    \includegraphics[width=0.50\textwidth]{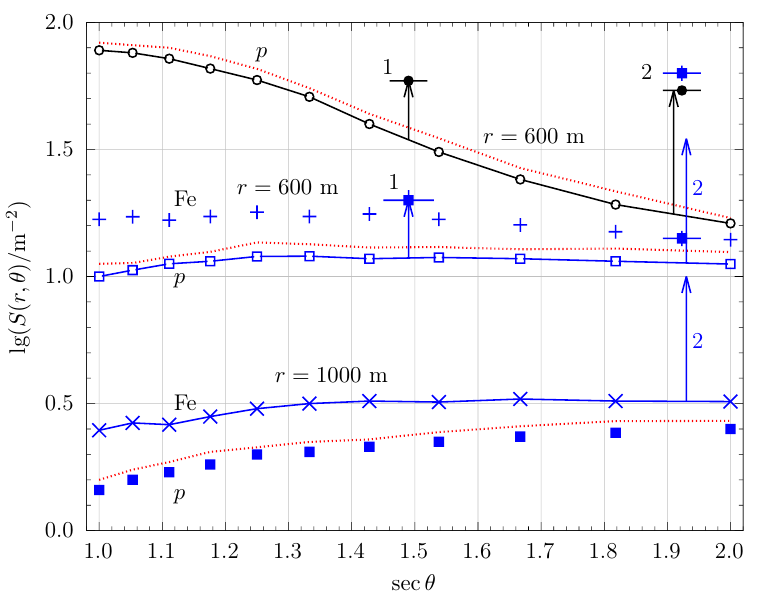}
    \caption{Fig.~1. Zenith-angular dependencies of responses in SD (empty circles) and MD with $1.0 \times \sec\theta$~GeV threshold at 600~m (empty squares, crosses) and 1000~m (dark squares, x-es) from the axis in showers with energy $E = 10^{19.5}$~eV obtained within the framework of the \qgsii{} model for primary protons ($p$) anf iron nuclei (Fe). Simulation results obtained using the \eposlhc{} model (dotted lines) were obtained for primary protons.}
    \label{f:1}
\end{figure}

The sequence of calculations for determining the $\SmuSim(E, \theta, r)$ value with relation \eqn{eq:13} is demonstrated in \fign{f:1} for a particular registered shower with energy $E = 10^{19.73}$~eV, which was estimated according to formulas \eqn{eq:4}, \eqn{eq:5}, \eqn{eq:7}, and zenith angle $\theta = 48\degr (\sec\theta = 1.49 \pm 0.03)$. For this event the value $\lg[\SsExp(E, 48\degr)] = 1.77 \pm 0.02$ was obtained (dark circle ``1''). The estimated energy of this event isn't used further and is only given for information. The value of parameter $\lg[\SsSim(E^*, 48\degr)] = 1.54$ for relation \eqn{eq:13} follows from the zenith-angular dependency displayed with empty circles. The length of the arrow ``1'' equals to the normalization shift:

\begin{align}
    &\Delta\lg[\Ssoo(48\degr)] = \nonumber \\
    &\qquad = \lg[\SsExp(E, 48\degr)] - \lg[\SsSim(E^*, 48\degr)] = \nonumber \\
    &\qquad\qquad = 1.77 - 1.54 = 0.23\text{,} \nonumber
\end{align}
which is to be added to the calculated value $\lg[\SmusSim(E^*, 48\degr)] = 1.08$ (curve with squares at $r = 600$~m) in order to obtain the expected density of the MD response:

\begin{align}
    &\lg[\SmusSim(E, 48\degr)] = \nonumber \\
    &\qquad = 1.08 + 0.23 = 1.31 \pm 0.02\text{.}
    \label{eq:14}
\end{align}
It appears that, within measurement errors, the value \eqn{eq:14} is close to the experimentally obtained density (dark square ``1''):

\begin{equation}
    \lg[\SmusExp(E, 48\degr)] = 1.30 \pm 0.04\text{.}
    \label{eq:15}
\end{equation}

Note that this result does not depend on the chosen hadron interaction model. To make it sure, let's consider, for example, zenith-angular characteristics of SD and MD responses obtained with the above mentioned technique using the \eposlhc{} model~\cite{b:20}. It is evident from \fign{f:1} that both models give curves with identical shapes and differing in absolute values by factor 1.1. Let's determine the parameter $\lg[\SmusSim(E^*, 48\degr)] = 1.60$ for relation \eqn{eq:13}. The length of arrow ``1'' in this case is
\begin{align}
    &\Delta\lg[\Ssoo(48\degr)] = \nonumber \\
    &\qquad = \lg[\SsExp(E, 48\degr)] - \lg[\SsSim(E^*, 48\degr)] = \nonumber \\
    &\qquad\qquad = 1.77 - 1.59 = 0.18\text{,} \nonumber
\end{align}
which is to be added to the new calculated value $\lg[\SmusSim(E, 48\degr)] = 1.12$ (dotted line at $r = 600$~m) in order to obtain the expected MD response density:

\begin{align}
    &\lg[\SmusSim(E, 48\degr)] = \nonumber \\
    &\qquad = 1.12 + 0.18 = 1.30 \pm 0.02\text{.}
    \label{eq:16}
\end{align}
It coincides with the value \eqn{eq:15}. From the considered example of two different hadron interaction models, with a certain degree of confidence one can assume that the considered shower was initiated by primary proton.

The equality of values \eqn{eq:14} and \eqn{eq:16} is not a random coincidence, but a natural characteristic of EAS development. In~\cite{b:21} it was shown that estimations of the CR mass compositions obtained from muon fraction with the use of expressions \eqn{eq:2} and \eqn{eq:3} within frameworks of different models were virtually identical. This feature of the behaviour of muon fraction is utilized in expression \eqn{eq:13}, where it is clearly seen in the identically transformed expression:

\begin{equation}
    \SmuSim(E, \theta, r) = \SsExp(E, \theta) \cdot
    \frac{
        \SmuSim(E^*, \theta, r) / \varepsilon
    }{
        \SsSim(E^*, \theta) / \varepsilon
    }
    \label{eq:17}
\end{equation}
with $\varepsilon \approx 1.1$ for \eposlhc. This coefficient varies within $(8-10)$\% in different hadron interaction models. From a physical standpoint, a formal substitution $\varepsilon \rightarrow E^*$ means normalization of all responses in \eqn{eq:17} by primary energy.

\subsection{The giant event}

\begin{figure}[!htb]
    \centering
    \includegraphics[width=0.50\textwidth]{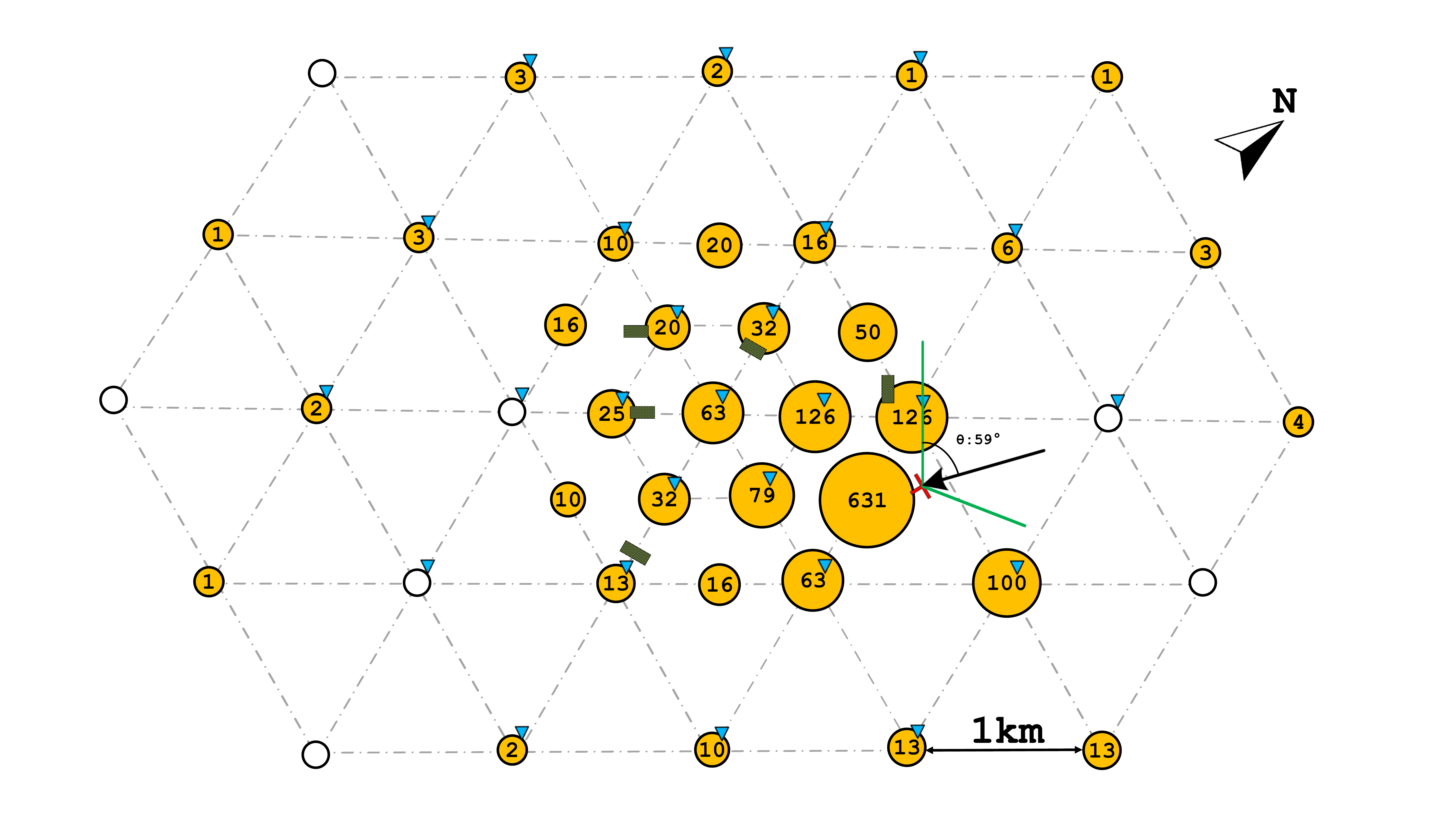}
    \caption{Fig.~2. The unique EAS event with energy $E \approx 1.12 \times 10^{20}$~eV on the array map: digits~--- summary SD responses; empty circles~--- stations that were not operating during the observation; rectangles~-- MDs with $1.0 \times \sec\theta$~GeV threshold; triangles~--- CLE detectors; cross and arrow indicate shower axis and arrival direction.}
    \label{f:2}
\end{figure}

\begin{figure}[!htb]
    \centering
    \includegraphics[width=0.50\textwidth]{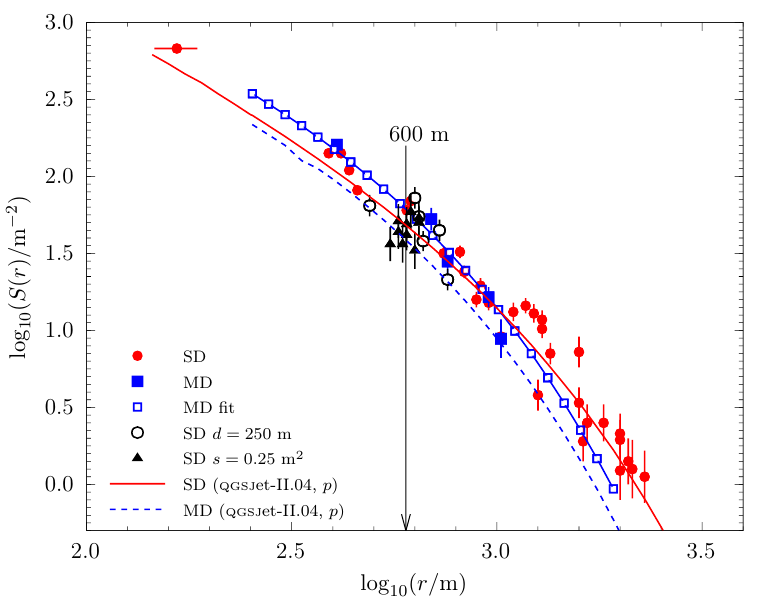}
    \caption{Fig.~3. Lateral distribution of particles in the Arian event registered in experiment by SDs (circles, triangles) and by MDs with $1.0 \times \sec\theta$~GeV threshold (dark squares). Solid and dotted lines represent theoretical LDFs of SD and MD responses calculated within the framework of the \qgsii{} model for primary protons with energy $10^{20}$~eV and $\theta = 60\degr$. Open squares and corresponding curve represent approximation \eqn{eq:12} of experimental data.}
    \label{f:3}
\end{figure}

On May 7 1989 a unique event was registered at the Yakutsk array, with energy $E \approx 1.12 \times 10^{20}$~eV and zenith angle $\theta = 58.7\degr \pm 1.3\degr$, and to this day it remains the most powerful event in its data bank. The shower was dubbed {\sl Arian}. In \fign{f:2} the shower footprint is shown on the array map with its axis, which was located with $\Delta r = 26$~m accuracy. The fired detector stations are shown with filled circles, their size is proportional to the logarithm of the number of measured SD responses. This shower covered the whole array and provided the opportunity to test its operation. In \fign{f:3} readings of all fired SD and MD are shown. Empty circles indicate measurements with six additional stations concentrated around the array center with 250~m spacing. Each station contained a single 2-\sqrm{} scintillation detector. Also during this period, a compact cluster of $0.25$-\sqrm{} scintillation detectors spaced by 50~m operated in the central area of the array. Readings of these detectors are shown with black triangles. All the data provided a precise measurement of the value $\lg[\SsExp(E, 59\degr)] = 1.735 \pm 0.013$ (dark circle ``2'' in \fign{f:1}). Solid curve indicates best fit of the SD data with approximation \eqn{eq:11}. It happened to be very close, in terms of both absolute value and shape, to the LDF obtained using the \qgsii{} model for primary protons with energy $10^{20}$~eV and $\theta = 60\degr$. Empty squares represent approximation \eqn{eq:12} of the experimentally measured lateral distribution of muons with energies $E_{\mu} \ge 1.92$~GeV. According to the $\chi^2$-test it satisfies the experimental data with parameters $b_{\mu} \approx 0.28$ and

\begin{equation}
    \lg[\SmusExp(E, 59\degr)] = 1.81 \pm 0.04\text{.}
    \label{eq:18}
\end{equation}
The density \eqn{eq:18} is represented in \fign{f:1} with dark square ``2''. Let's determine the normalization shift for this shower according to the procedure described in Section~2.4:

\begin{align}
    & \Delta\lg[\Ssoo(59\degr)] = \nonumber \\
    & \quad = \lg[\SsExp(E, 59\degr)] - \lg[\SsSim(E^*, 59\degr)] = \nonumber \\
    & \quad\quad = 1.735 - 1.225 = 0.5\text{,}
    \label{eq:19}
\end{align}
which is shown with arrow ``2'' in \fign{f:1} and calculate the expected density of muon response:

\begin{align}
    &\lg[\SmusSim(E, 59\degr, p)] = \nonumber \\
    &\qquad = 1.04 + 0.5 = 1.54 \pm 0.02\text{,}
    \label{eq:20}
\end{align}
which turned out to be not only lower than the expected value from primary protons by $\Delta_{\mu} = 1.81 - 1.54 = 0.27$, but also lower than the expected value from iron nuclei:

\begin{align}
    &\lg[\SmusSim(E, 59\degr, \text{Fe})] = \nonumber \\
    &\qquad = 1.16 + 0.5 = 1.66 \pm 0.02
    \label{eq:21}
\end{align}
by $\Delta_{\mu} = 1.81 - 1.66 = 0.15$. Similar result follows from muon density at axis distance $r = 1000$~m, where the expected value for iron nuclei is:

\begin{align}
    &\lg[\SmutSim(E, 59\degr, \text{Fe})] = \nonumber \\
    &\qquad = 0.51 \pm 0.5 = 1.01 \pm 0.02\text{,}
    \label{eq:22}
\end{align}
and the measured value is

\begin{equation}
    \lg[\SmutExp(E, 59\degr)] = 1.15 \pm 0.04\text{,}
    \label{eq:23}
\end{equation}
shown in \fign{f:1} with dark square.

The relation \eqn{eq:13} is convenient for implementing this method because it eliminates the need to perform heavy and time consuming Monte Carlo calculations of $\SmuSim(E, \theta, r)$ in each individual shower. In this case we have performed them only for showers with $E^* = 10^{19.5}$~eV and zenith angle values separated by step $\Delta\cos\theta = 0.1$. This approach is fundamentally different from that used at the TA and Auger arrays, where each registered EAS event is simulated with Monte Carlo method $\sim 10^6$ times to estimate its main parameters. The previously performed computational analysis of the EAS development have demonstrated that estimations of the CR mass composition obtained with the use of the relation \eqn{eq:13} are quite correct. This is evident from the above-mentioned example of the giant shower where experimentally measured parameters were $\theta = 58.7\degr \pm 1.3\degr$ and $\lg\SsExp(E, 59\degr)= 1.735 \pm 0.013$. Densities of the MD response calculated for this event with Monte-Carlo method using the \qgsii{} model for primary proton are

\begin{equation}
    \lg[\SmusSim(E, 59\degr, p)] = 1.65 \pm 0.02\text{,}
    \label{eq:24}
\end{equation}
and

\begin{equation}
    \lg[\SmutSim(E, 59\degr, p)] = 0.89 \pm 0.02\text{,}
    \label{eq:25}
\end{equation}
which within errors are close to the expected values from the relation \eqn{eq:13} and \fign{f:1}.

\section{Results and discussion}

\subsection{Correlation dependencies}

The geometry of the Yakutsk array during the observational period 1974.01.01 - 1990.06.23 is shown in \fign{f:2}. In 1976 two new 36-\sqrm{} MDs were deployed at $\approx 500$~m from the array center. In 1986 another three 20-\sqrm{} MDs were deployed. During the renovation in 1990-1992, the outermost SD stations were dismantled and moved to a central circle with a radius 2~km. The current analysis of the MD data includes only showers with axes lying inside this circle.

\begin{figure}[!ht]
    \centering
    \includegraphics[width=0.50\textwidth]{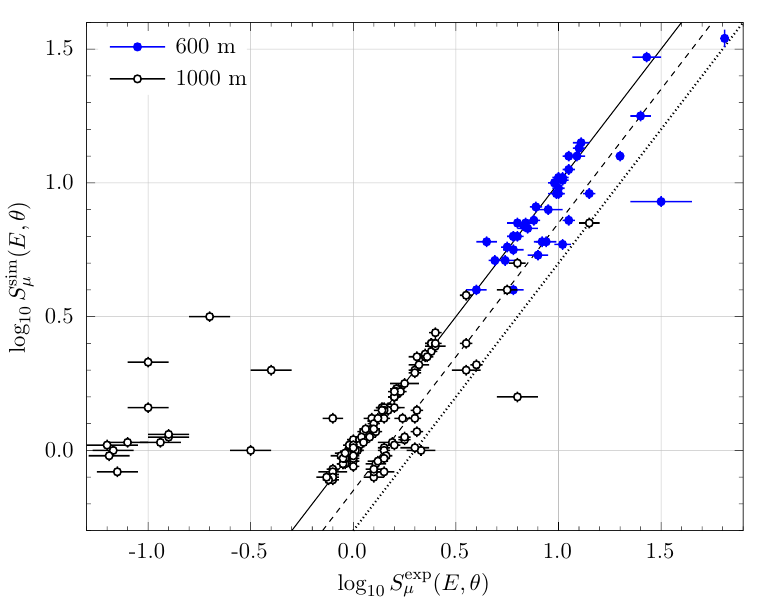}
    \caption{Fig.~4. Correlations of the response densities recorded by MDs with $1 \times \sec\theta$~GeV threshold at axis distances 600~m (dark circles) and 1000~m (open circles) in 127 showers with energies $E \ge 10^{19.1}$~eV and $\theta \le 60\degr$. Horizontal axis~--- densities determined in experiment, vertical axis~--- those calculated from simulation results performed within the framework of \qgsii{} model for primary protons (solid line), iron nuclei (dashed line). Dotted line highlights EASs with abnormally high muon content.}
    \label{f:4}
\end{figure}

In \fign{f:4} a sample of 127 showers with $E \ge 10^{19.1}$~eV and $\theta \le 60\degr$ is shown. It includes events with muons registered at axis distances $600$~m (dark circles) and $1000$~m (open circles). Showers with at least two fired MDs were selected for the analysis. A detector was considered fired even if it didn't register a single particle but was in standard accepting mode. Lines represent the expected correlations between measured and expected muon densities in EASs initiated by different primary particles. Solid line corresponds to the correlation for primary protons. Dashed line indicates correlation of muon densities in EASs originating from iron nuclei. It was obtained by averaging the differences of muon densities shown in \fign{f:1} in all interval of zenith angles. One can see that it is shifted to the right along the horizontal axis from the proton correlation line by the value

\begin{equation}
    \mean{\Delta_{p-\text{Fe}}} = \lg\frac{
        \SmuSim(E, \theta, r, \text{Fe})
    }{
        \SmuSim(E, \theta, r, p)
    } \approx 0.15\text{.}
    \label{eq:26}
\end{equation}

The dotted line in \fign{f:4} indicatively highlights a few showers (further marked with the symbol ``X'') with abnormally high muon content. It is shifted to the right along the horizontal axis from the proton correlation line by the value

\begin{equation}
    \mean{\Delta_{p-\text{X}}} = 2 \times \mean{p-\text{Fe}} \approx 0.30\text{.}
    \label{eq:27}
\end{equation}

The considered sample includes showers with at least 2 MDs fired within the axis distance range $800-1600$~m. Their readings were approximated with LDF \eqn{eq:12} and the densities $\SmutExp(E,\theta)$ were determined. In this case, the steepness parameter $\beta_{\mu}(\theta)$ was taken equal to the mean value following from the previously obtained experimental data~\cite{b:22}, as it was impossible to calculate it in the majority of individual events. About 27\% of all events had one MD with readings (or more, in rare cases) at axis distances from 300~m to 800~m. In such a case, the $\SmusExp(E,\theta)$ density was determined in addition to $\SmutExp(E,\theta)$. It was the most optimal approach to data selection which had its impact on the distribution of some events presented in \fign{f:4}.

\begin{figure}[!htb]
    \centering
    \includegraphics[width=0.50\textwidth]{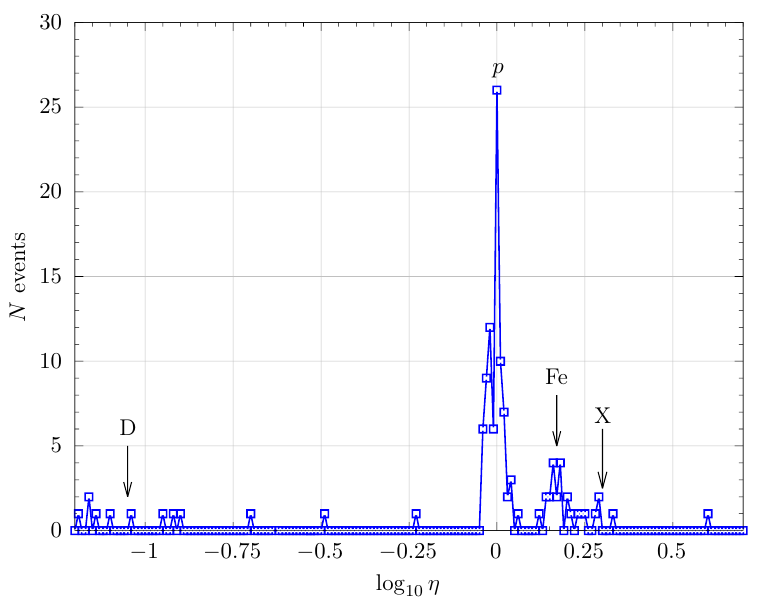}
    \caption{Fig.~5. Deviations of individual pairs of muon densities from correlation lines displayed in \fign{f:4} at axis distance 1000~m.}
    \label{f:5}
\end{figure}

Deviations of individual pairs of muon densities from the correlation lines presented in \fign{f:4} are shown in \fign{f:5}:

\begin{equation}
    \eta = \SmuExp / \SmuSim(p)
    \label{eq:28}
\end{equation}
Here zero-value on horizontal axis corresponds to the position of proton peak according to the \qgsii{} model. Labels ``Fe'' and ``X'' indicate shifts \eqn{eq:26} and \eqn{eq:27} relative to this peak. Events marked with the symbol ``D'' refer to events with observed muon deficit. They are positioned at peaks with mean values in the region of their location at 68\% confidence level. For all considered groups of events the following values were obtained for deviations \eqn{eq:28}:

\begin{gather}
    \eta(p) = 0.99_{-0.08}^{+0.08}\text{,}\label{eq:29} \\
    \eta(\text{Fe}) = 1.44_{-0.12}^{+0.18}\text{,}\label{eq:30} \\
    \eta(\text{X}) = 1.95_{-0.33}^{+2.05}\text{,}\label{eq:31} \\
    \eta(\text{D}) = 0.09_{-0.02}^{+0.05}\text{.}\label{eq:32}
\end{gather}
Thus, the distribution of peaks presented in \fign{f:5} can be interpreted as a probable composition of primary particles.

\begin{table*}[!ht]
    \centering
    \caption{Table~1. The number of EAS events in energy bins with a width $\Delta\log_{10}E=0.1$.}
    \label{t:1}
    \begin{tabular}{l|rr|rr|rr|rr|r}
        \hline
        \hline
        $\mean{\log_{10}E}$ &
        $p$ & $w_p$ & Fe & $w_{\rm Fe}$ & X & $w_{\rm X}$ &
        $\gamma$ & $w_{\gamma}$ & Total \\
        \hline
        20.05 & 0 & 0 & 0 & 0 & 1 & 1 & 0 & 0 & 1 \\
        19.75 & 1 & 0.5 & 1 & 0 & 1 & 0.5 & 0 & 0 & 2 \\
        19.65 & 4 & 1 & 0 & 0 & 0 & 0 & 0 & 0 & 4 \\
        19.55 & 9 & 0.75 & 1 & 0.09 & 0 & 0 & 2 & 0.16 & 12 \\
        19.45 & 11 & 0.73 & 1 & 0.07 & 2 & 0.13 & 1 & 0.07 & 15 \\
        19.35 & 16 & 0.76 & 3 & 0.14 & 0 & 0 & 2 & 0.1 & 21 \\
        19.25 & 27 & 0.71 & 7 & 0.18 & 0 & 0 & 4 & 0.11 & 38 \\
        19.15 & 20 & 0.59 & 8 & 0.24 & 2 & 0.06 & 4 & 0.12 & 34 \\
        \hline
        Total & 88 & 0.69 & 21 & 0.17 & 5 & 0.04 & 13 & 0.1 & 127 \\
        \hline
        \hline
    \end{tabular}
\end{table*}

\subsection{CR composition}

The differentiated data on the CR composition depending on the EAS energy are given in Table~\ref{t:1}. It is evident that the fraction of protons in this sample is nearly constant and amounts to $\approx 69$\% except for the last row. The fraction of showers with muon deficit is also roughly constant and is about 10\%. It does not contradict our previous estimation~\cite{b:23, b:24}. The fraction of iron nuclei gradually increases from 0 to 24\%. In \cite{b:23} this fraction was estimated to be about 36\% from the sample of 33 events. The difference between the previous results and Table~\ref{t:1} probably arises from different sizes of the considered samples. It is also possible that different methods of estimation of the CR composition are also reflected here.

There are 5 showers worth noting in the column ``X'' with the observed abnormally high muon content. A detailed analysis of the obtained data have confirmed that these were the real densities recorded by MDs. From lateral distribution of particles in the most powerful event presented in \fign{f:3} it is evident that the signals recorded by SDs and MDs were close in value and probably were produced by muons with energies $E_{\mu} \ge 1.92$~GeV. At axis distances $r \ge 400$~m the contribution from electro-magnetic component is almost absent. So far it is hard to interpret this fact in any way.

The positions of peaks remained virtually unchanged relative to each other. This indicates that {\sl the muon correlation method} is not sensitive to the choice of hadron interaction model.

Let's estimate the parameter $\mean{\zA}$ \eqn{eq:3} from the data in Table~\ref{t:1}. For this we will use the expression:

\begin{align}
    \ln\mean{\rhoMu{\text{exp.}}} - \ln{\rhoMu{p}} & =
    2.3 \times \lg\left(
        \frac{\SmuExp}{\SmuSim(p)}
    \right) = \nonumber \\
    & \quad = -0.127 \pm 0.035\text{,}
    \label{eq:33}
\end{align}
where the right side can be determined from the data in \fign{f:4} and \fign{f:5} by summation of all 127 events. For this sample with average energy $\mean{E} \approx 1.82 \times 10^{19}$~eV and $\mean{\sec\theta} \approx 1.33$ we will obtain the following value:

\begin{equation}
    \mean{\zA} = \frac{-0.127}{\ln\rhoMu{\text{Fe}} - \ln\rhoMu{p}} = 
    -0.03 \pm 0.02\text{,}
    \label{eq:34}
\end{equation}
which within errors agrees with our earlier estimations of this parameter obtained from mean LFDs~\cite{b:6, b:7, b:8}. Formally, one can conclude from \eqn{eq:34} and formula \eqn{eq:2} that in the considered energy range CRs consist mainly of protons. But in fact this is not the case. Estimations of nuclear composition of primary particles derived from the EAS muon content, on average, can be significantly distorted due to certain fraction of muon-poor and muon-rich EAS events presented in the data.

\section{Conclusion}

A new algorithm for estimation of the CR mass composition in individual EAS events is described above. It is based on a comparison of the measured MD responses and those calculated using the \qgsii{} hadron interaction model for primary particles with a given mass. In this particular case, protons with zenith angles $\theta \le 60\degr$ were considered. The chosen model accurately describes the development of all EAS components. It has proven to be highly effective during the investigation of the so-called ``muon puzzle''~\cite{b:7, b:8, b:21}. Calculations of SD and MD responses for this work were performed without resorting to the Monte Carlo method for individual events. For this purpose, zenith-angular dependences of these values were obtained beforehand (\fign{f:1}). This approach significantly simplified the analysis of the air showers muon content without losing the quality of the obtained results.

Comparison of experimentally determined and calculated responses for all EASs was made for a given value of CR energy $E^* = 3.16 \times 10^{19}$~eV. The measured MD responses in individual showers with energy $E$ were normalized by the corresponding calculated values according to relation \eqn{eq:13}. The essence of the method is demonstrated in \fign{f:4}, where correlations of the MD response densities at axis distances $r = 600$~m and 1000~m are shown (such picture is observed at any axis distance). We selected events with densities determined at $1000$~m (as in~\cite{b:23, b:24}), which give maximum statistics with a good quality of muon data. The relation between the experimentally derived muon response and the expected from the \qgsii{} model is shown in \fign{f:5}. This distribution has several pronounced peaks. The first one \eqn{eq:29} undoubtedly refers to protons. The second peak \eqn{eq:30} conventionally could be attributed to iron nuclei. Third one \eqn{eq:31} is formed by several events with abnormally high muon content, including the largest air shower registered at the Yakutsk array (see \fign{f:2} and \fign{f:3}). Its origin is yet to be discovered. The fourth peak \eqn{eq:32} refers to abnormally muon-poor EAS events. Probably it has direct connection to primary gamma photons. Some characteristics of events in these peaks are presented in Table~\ref{t:1}.

Further plans are to continue studying the CR mass composition by applying this method to the Yakutsk EAS array data in the lower energy range.

\section*{Acknowledgements}

This work was made using the data obtained at The Unique Scientific Facility ``The D.\,D.~Krasilnikov Yakutsk Complex EAS Array'' (YEASA) (\url{https://ckp-rf.ru/catalog/usu/73611/}). Authors express their gratitude to the staff of the Separate structural unit YEASA of ShICRA SB RAS.

\section*{Funding}

This work was made within the framework of the state assignment \#\,122011800084-7.

\section*{Conflict of interest}

The authors of this work declare that they have no conflict of interest.

\end{document}